\documentclass[a4paper,11pt,twoside]{article}

\usepackage{graphics}

\usepackage{amsmath,amssymb,amsfonts}
\usepackage{graphicx}
\usepackage{authblk}
\usepackage{multicol}
\usepackage[hmargin=2cm,vmargin=1.5cm]{geometry}
\usepackage{float}
\usepackage[font=it]{caption}
\usepackage[english]{babel}
\usepackage[scaled]{helvet}
\usepackage[T1]{fontenc}
\usepackage{url}
\usepackage[colorlinks=true, allcolors=blue]{hyperref}
\usepackage[numbers,sort&compress]{natbib}
\makeatletter
\renewcommand\@biblabel[1]{ #1.}

\def\tagform@#1{\maketag@@@{\bfseries(\ignorespaces#1\unskip\@@italiccorr)}}
\renewcommand{\eqref}[1]{\textup{{\normalfont Eq.~(\ref{#1}}\normalfont)}}
\makeatother

\title{Multi-line fiber laser system for cesium and rubidium atom interferometry}

\author[1]{Cl\'ement Diboune}
\author[1]{Nassim Zahzam}
\author[1]{Yannick Bidel}
\author[1,2]{Malo Cadoret}
\author[1]{Alexandre Bresson}
\affil[1]{ONERA - The French Aerospace Lab, F-91761 Palaiseau, France}
\affil[2]{Laboratoire Commun de M\'etrologie CNAM, 61 rue du Landy,
93210 La Plaine Saint-Denis, France}

\date{}
\begin{document}
\maketitle

\begin{abstract}
We present an innovative multi-line fiber laser system for both cesium and rubidium manipulation. The architecture is based on frequency conversion of two lasers at 1560 nm and 1878 nm. By taking advantage of existing high performance fibered components at these wavelengths, we have demonstrated multi-line operation of an all fiber laser system delivering 350 mW at 780 nm for rubidium and 210 mW at 852 nm for cesium. This result highlights the promising nature of such laser system especially for Cs manipulation for which no fiber laser system has been reported. It offers new perspectives for the development of atomic instruments dedicated to onboard applications and opens the way to a new generation of atom interferometers involving three atomic species (\textsuperscript{85}Rb, \textsuperscript{87}Rb and \textsuperscript{133}Cs) for which we propose an original laser architecture.
\end{abstract}

\begin{multicols}{2}

\section{Introduction}
Atom interferometry has shown to be a powerful tool for precision measurements especially through the demonstration of high performance inertial sensors like gravimeters \cite{Peters:01,Muller:08,Hu:13,Gillot:14}, gradiometers \cite{McGuirk:02,Asenbaum:16} or gyroscopes \cite{Gustavson:97,Dutta:16}. These instruments appear as promising candidates for testing fundamental physics like detecting gravitational waves \cite{Dimopoulos:08}, exploring short-range forces \cite{Wolf:07} or testing the Universality of Free Fall (UFF) \cite{Altschul:15}. The last point is particularly under investigation with the development of atom interferometer experiments interrogating simultaneously two atomic species \cite{Bonnin:13, Zhou:15, Schlippert:14, Barrett:16}. In addition to testing the UFF, multi-species atom interferometers allow new architectures of cold atom inertial sensors \cite{Bonnin:161,Bonnin:162,Johnson:16} which could reduce systematic effects, extend their dynamic range, perform simultaneously multi-axis measurements or eliminate dead time measurements. Cesium and rubidium atoms seems a very promising couple for multi-species atom interferometry. Indeed, these two species have demonstrated the best performances for inertial sensors \cite{Peters:01,Muller:08,Hu:13,Gillot:14,McGuirk:02,Asenbaum:16,Gustavson:97,Dutta:16}.  However, the laser wavelengths used to manipulate rubidium and cesium are significantly different (780 nm for Rb and 852 nm for Cs) conducting to a rather complex free space laser system difficult to implement in an onboard inertial sensor or for a space test of the UFF. The development of a compact and robust laser system for manipulating cesium and rubidium atoms is thus particularly interesting for these demanding onboard applications but also more globally for laboratory experiments manipulating Rb and Cs atoms like those aiming to test the local position invariance using \textsuperscript{87}Rb and \textsuperscript{133}Cs fountains \cite{Guena:12} or to study ultracold heteronuclear molecules \cite{Molony:14}. More specifically, cesium atoms are extensively used in a wide range of applications whether for commercial or research purposes, for instance for high performance magnetometry, atom lithography for nanofabrication, or in atomic clocks used as a primary frequency standard. For such applications, a compact and reliable laser system at 852 nm is of prime importance and an all-fibered laser system could open new promising perspectives.

Important developments have been conducted to achieve compact and robust laser systems for rubidium atoms based either on solid-state laser diodes emitting directly at 780 nm \cite{schkolnik:16} or on Second Harmonic Generation (SHG) of a telecom fiber bench \cite{Carraz:09}. The main advantages of this second solution is that the laser system relies on the great maturity of fiber components in the telecom C-band, many of them meeting Telcordia standards of qualification, reducing consequently the amount of free-space optics and making the setup more compact, reliable and less sensitive to misalignment. Concerning power issues, solid-state laser diodes are usually operated with tapered amplifiers offering up to typically 3.5 W but suffering from significant sensitivity to optical feedback, from aging and providing a relatively poor output beam quality factor which could impact for instance further fiber coupling efficiency. On the other hand, commercial Erbium-Doped Fiber Amplifiers (EDFA) combined with fiber lasers or pigtailed laser diodes at 1.5 $\mu$m are readily available and have demonstrated a high level of reliability (telecordia standards, space qualifications), and yielding a very low spectral linewidth (< 1 kHz) which is hardly achievable with common solid-state technology at 780 nm. The use of commercial EDFA's at 1.5 $\mu$m offers also a broader range of available power, reaching even more than 1 kW for some applications. A huge interest of working at the telecom wavelength comes also from the availability of high quality fibered optical components such as fibered AOM's for power control or fibered phase/intensity modulators for generating easily all the required laser lines for typical atom interferometer experiment, allowing the use of only a single fiber laser source \cite{Theron:15}.

For cesium, at 852 nm, no fiber laser system have been yet demonstrated and up to now only free-space solid-state solutions have been implemented. The most advanced development was done for the PHARAO space cold atom clock project which led to a space qualified free space laser system \cite{Leveque:15} at the expense of a very complex architecture to ensure compactness and stability against misalignment.

Here, we present a multi-line fiber laser system addressing both rubidium and cesium atoms based on the frequency conversion of lasers at 1560 nm and 1878 nm. The laser at 780 nm for Rb atoms is obtained by frequency doubling a telecom laser at 1560 nm. This method is now well known and leads to compact and robust fiber laser systems. This kind of laser system is now implemented on many atom interferometer experiments and has shown its reliability for onboard applications \cite{Geiger:11,Bidel:13}. The laser at 852 nm is based on Sum-Frequency Generation (SFG) of the laser at 1560 nm and a laser at 1878 nm. To our knowledge, this method has never been proposed and should lead to a reliable and compact fiber system. It takes advantage of the telecom fiber technology at 1.5 $\mu$m and of the fiber technology at 2 $\mu$m which are currently in intensive development for many applications \cite{Geng:14} like surgery, material processing or LIDAR. For instance commercial Thulium-Doped Fiber Amplifiers (TDFA) allow to reach optical power exceeding 1 kW and laser sources with spectral linewidth below 1 kHz are readily available. We want to highlight here that this original architecture to produce 852 nm still allows the use of telecom optical components at 1.5 $\mu$m such as phase/intensity modulators or fibered AOM's to generate additional laser lines at 852 nm or to control the laser power output. 

In the following, we give a description of our tested laser system followed by its characterization. In the last part, we propose an original laser architecture very promising for  \textsuperscript{85}Rb, \textsuperscript{87}Rb and \textsuperscript{133}Cs atom interferometry. 

\section{Laser system description}
\begin{figure}[H]
\centering
\includegraphics[width=8cm]{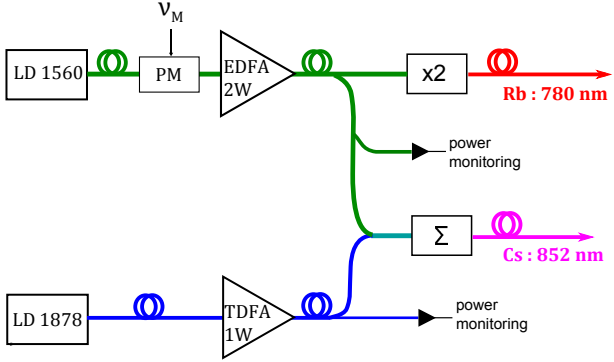}
\caption{Scheme of the tested laser system producing light at 780 nm for Rb and 852 nm for Cs (LD, laser diode; PM, phase modulator; $\nu_M$ is the microwave modulation frequency fed to the PM; x2, second harmonic generation in a PPLN waveguide crystal; $\Sigma$, sum-frequency generation in a PPLN waveguide crystal).}
\label{laser}
\end{figure}

The laser system that we have tested is described on Fig. \ref{laser} and is only composed of fiber components. The laser source at 1560 nm is a DFB laser diode (output power: 29 mW, linewidth: 0.1 MHz) amplified with a 2 W EDFA. A phase modulator is inserted between the laser source and the EDFA to demonstrate the ability to generate additional laser lines necessary for instance during both atom cooling and interferometry phases. This is a crucial point since it allows to reduce significantly the number of needed laser sources and optical amplifiers, and at the same time the complexity and size of the complete laser system \cite{Bonnin:161,Theron:15}. The laser source at 1878 nm is a DFB laser diode (output power: 3 mW, linewidth < 2 MHz) amplified with a 1 W TDFA. The laser at 780 nm is obtained by SHG of the laser at 1560 nm in a periodically poled lithium niobate (PPLN) waveguide crystal. An output power of 350 mW at 780 nm is achieved with an input power of 700 mW at 1560 nm. The laser at 852 nm is obtained by combining the laser at 1560 nm and the laser at 1878 nm thanks to a fiber coupler and then by achieving SFG in a PPLN waveguide crystal. A power of 210 mW at 852 nm is obtained with an input power of 300 mW at 1560 nm and 300 mW at 1878 nm. If we dedicate our experimental setup only to produce 852 nm for cesium manipulation, and if we couple directly the 2 W EDFA and 1 W TDFA into the PPLN waveguide, we estimate by extrapolation that we could achieve a maximum output power of roughly 430 mW. Note that the demonstrated maximum output powers at both 780 nm and 852 nm are not a fundamental limit, and could be increased for higher input powers, limited in our case by the maximum power of the TDFA. Using the same kind of fibered optical components and SHG, L\'ev\`eque \textit{et al.}  have demonstrated for instance output power reaching 1 W at 780 nm for 1.8 W input power \cite{Leveque:14}. Moreover, implementing our proposed solution to generate 852 nm in a free space configuration should lead to even higher laser power. For instance, more than 11 W optical power has been obtained at 780 nm using SHG in free space \cite{Sane:12}.

\section{Laser system characterization}
\begin{figure}[H]
\centering
\includegraphics[width=8cm]{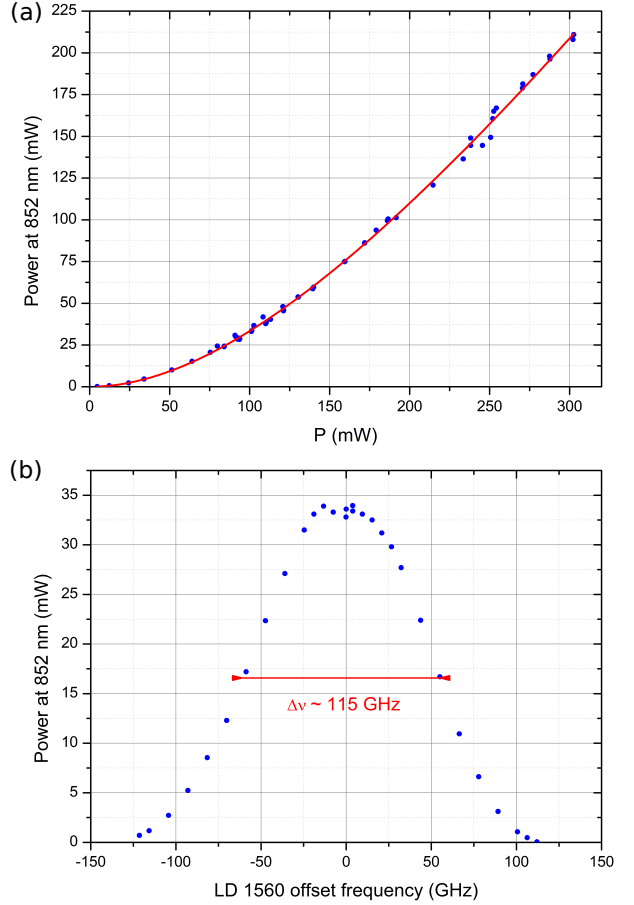}
\caption{(a) Efficiency of the sum-frequency generation with $P=P_{1560}=P_{1878}$. The blue points are experimental data. The red curve is the fit with Eq. (\ref{conv}).  (b) Frequency dependance of the sum-frequency generation by tuning the 1560 nm laser diode.}
\label{rendement}
\end{figure}

The 852 nm generation part of this scheme was precisely characterized in terms of sum-frequency generation and spectral properties. We do not report here the characterization of the 780 nm laser which was already investigated in other articles \cite{Leveque:14}. First, we estimated the conversion efficiency by measuring the output power at 852 nm versus the input power at 1560 nm and 1878 nm. In this measurement, the input power at 1560 nm and 1878 nm were kept equal and the temperature of the crystal was optimized to 52$^{\circ} \mathrm{C}$. We did not notice any significant variations of the optimum temperature versus the input power. The results obtained are shown in Fig. \ref{rendement}(a). The measurements were fitted by a theoretical expression [Eq. (\ref{conv})] derived from the one of SFG conversion efficiency in the approximation of equal number of photons for the two pump lasers \cite{Smith:16}:

\begin{equation}
P_{852}=\epsilon\cdot 2\,P\cdot\tanh^2\left(\sqrt{\frac{\eta\cdot P}{2\epsilon}}\right)
\label{conv}
\end{equation}

In this Eq., $P=P_{1560}=P_{1878}$ is the input power of each pump laser, $P_{852}$ is the output power of the SFG at 852 nm, $\epsilon$ represents the total loss coefficient in the SFG component and $\eta$ represents the effective conversion efficiency between the fiber input and the fiber output in the undepleted-pump approximation:

\begin{equation}
P_{852}\approx \eta\cdot P^2\;\;\;\text{,}\;\;\;\frac{\eta\cdot P}{2\epsilon}\ll 1
\label{conv2}
\end{equation}

The fit gives a conversion efficiency equal to $\eta = 419 \pm 19\; \%/W$ similar to the SHG conversion efficiency at 780 nm and a loss coefficient $\epsilon=0.57\pm 0.02 $.

The power stability and the polarization extinction ratio (PER) of our laser system were also characterized. At 780 nm, for an output power of 350 mW, a power stability of 0.2\% rms over 6 hours and a PER of 23 dB were obtained. At 852 nm, for an output power of 200 mW, a power stability of 1.8\% rms over 6 hours and a PER of 30 dB were obtained. By monitoring simultaneously the input powers in the SFG PPLN crystal at 1878 nm and 1560 nm and the output power at 852 nm, we deduced that the power stability at 852 nm is limited by the power fluctuation at 1878 nm at the output of the TDFA.

We also estimated the spectral acceptance of the SFG by measuring the power at 852 nm against the frequency of the laser at 1560 nm. The results are shown in Fig. \ref{rendement}(b). The full width at half maximum (FWHM) of this curve is equal to 115 GHz. This spectral acceptance is large enough to allow the generation of all the laser frequencies needed for a Cs atom interferometry experiment which are spread over 10 GHz. In terms of output power versus crystal temperature, one obtained a dependency with a FWHM of 2.5$^{\circ} \mathrm{C}$.

\begin{figure}[H]
\centering
\includegraphics[width=8cm]{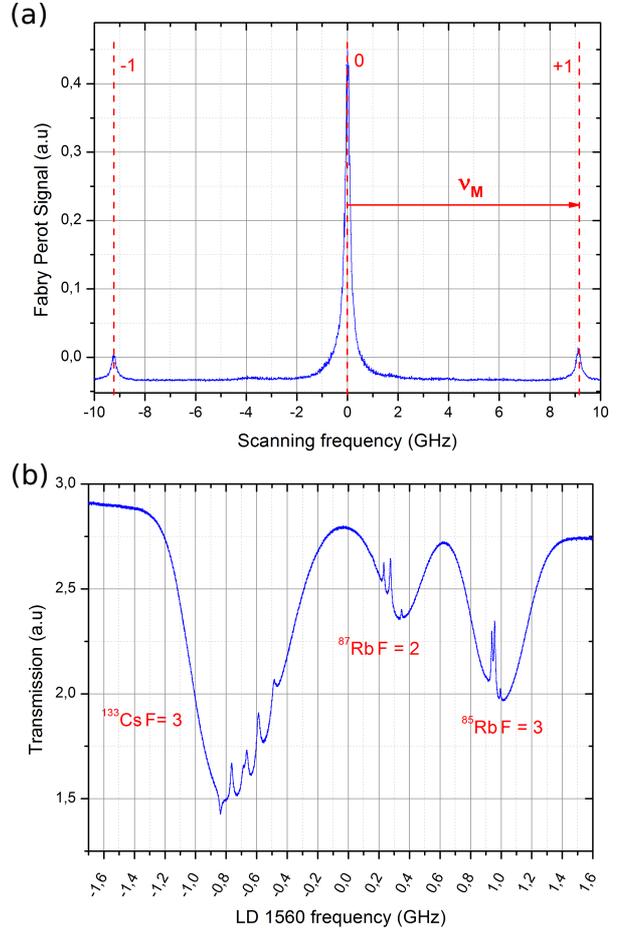}
\caption{(a) Fabry-Perot spectrum of the phase-modulated laser at 852 nm. The modulation is made at 1560 nm with a frequency $\nu_M \approx 9.2$ GHz. (b) Saturated absorption signal on Rb and Cs while scanning the frequency of the laser at 1560 nm. Note that on Fig. \ref{abssat}(b) the frequency separation between the Cs and Rb lines is adjustable by controlling the 1878 nm laser diode wavelength.}
\label{abssat}
\end{figure}

The ability of such laser system to generate simultaneously several laser lines is highlighted by Fig. \ref{abssat}(a). Microwave signal at $\nu_M \approx 9.2$ GHz is sent to the phase modulator in the 1.5 $\mu$m branch represented in Fig. \ref{laser}. This modulation frequency corresponds roughly to the frequency between the two hyperfine ground states of cesium. The carrier frequency (0 order modulation) and the first sideband (+1 order modulation) can be used as the cooling and repumping lines during magneto-optical trapping and as the two Raman laser lines during the interferometry phase. This method allows the laser system to be free from any phase lock loop between the two Raman lines, which otherwise is mandatory for light pulse atom interferometry. Injecting simply all the needed microwave modulation frequencies into the phase modulator would allow the laser system to cool and trap simultaneously the three atomic species \textsuperscript{133}Cs, \textsuperscript{87}Rb and \textsuperscript{85} Rb, which would have required otherwise at least 6 laser sources. Note that the use of phase or intensity modulation for Raman laser lines generation leads also to the production of parasitic lines that impact the interferometer output through AC Stark effect \cite{Carraz:12}. For most terrestrial onboard applications these parasitic laser lines have negligible impacts. However, for high precision measurements aiming for instance to test the UFF at a high level, some methods could be further implemented to reduce the impact of these parasitic laser lines \cite{Carraz:12} or a Raman phase lock could be implemented, keeping the phase/intensity modulation scheme only for cooling and trapping the atoms.

The spectral linewidth of the 852 nm laser was estimated by using a confocal scanning Fabry-Perot interferometer. The FWHM of the peaks obtained was equal to 5 MHz limited by the Fabry-Perot resolution. One can deduce that the spectral linewidth of our laser is below 5 MHz. This value is compatible with the expected spectral linewidth (< 2 MHz) from the datasheet of the laser diodes at 1560 nm (0.1 MHz) and at 1878 nm (< 2 MHz). The estimated linewidth is below the natural linewidth of the D2 transition of cesium (5.2 MHz) and allows to operate an atom interferometer experiment. In order to reach the maximum sensitivity of the atom interferometer \cite{LeGouet:07}, the spectral linewidth of the 852 nm laser could be improved by using a narrower laser source at 1878 nm like a whispering gallery mode laser diode \cite{Liang:15}, or a more standard thulium fiber laser.

We performed also a saturated absorption spectroscopy simultaneously on both rubidium and cesium. For this experiment, the two beams at 852 nm and 780 nm were recombined and sent into a classical saturated absorption setup containing two cells, one with rubidium vapor and the other with cesium vapor. The frequency sweep was done by acting only on the frequency of the 1560 nm laser diode giving simultaneously a frequency sweep on both laser lines at 852 nm and 780 nm. One can notice that in this configuration the amplitude of the frequency sweep is double at 780 nm compared to 852 nm. The simultaneous saturated absorption spectrum for \textsuperscript{133}Cs, \textsuperscript{87}Rb and \textsuperscript{85} Rb is shown on Fig. \ref{abssat}(b). This spectrum highlights the fact that our laser system allows to manipulate simultaneously these three atomic species.

\begin{figure}[H]
\centering
\includegraphics[width=8cm]{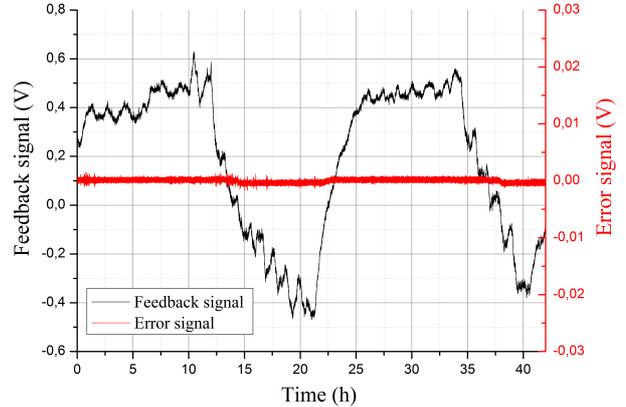}
\caption{Evolution of the feedback signal sent to 1.5 $\mu$m laser diode and the associated error signal coming from lock-in detection when the 852 nm is locked on a saturated absorption peak. The scale factor associated to the feedback signal according to frequency variation of the laser source is 95 MHz/V. The scale factor associated to the error signal according to frequency variation at 852 nm is estimated at approximately 50 kHz/mV.}
\label{laserlock}
\end{figure}

Finally, to evaluate the frequency long term stability of such fibered laser system, we have demonstrated the ability to lock the 852 nm output to a cesium transition. The 852 nm output beam was sent into a classical saturated absorption setup comprising a cesium vapor cell. For this demonstration, the 780 nm beam was not used. The 852 nm was locked to the F=3 - F'=2x3 saturated absorption peak using a standard lock-in detection technique by modulating the 1.5 $\mu$m laser diode current at 15 kHz. The feedback is applied to the 1.5 $\mu$m laser diode through a typical PID-control. This laser locking scheme underlines the possibility to lock the 852 nm by acting only on the 1.5 $\mu$m laser diode, compensating simultaneously the drift of both laser sources. Note that this locking scheme could have been implemented also by acting only on the 1878 nm laser diode. 
Fig. \ref{laserlock} shows the feedback signal sent to the 1.5 $\mu$m laser diode and the error signal coming from the lock-in detection when the laser at 852 nm is locked. The evolution of the feedback signal reveals the intrinsic frequency instability of both laser diodes which stays lower than 100 MHz over 40 hours. This frequency instability is efficiently compensated by the laser lock as shown on the error signal which exhibits a maximum signal deviation of 3 mV, much smaller than the 100 mV error signal variation that should correspond roughly to the atomic linewidth of 5.2 MHz.

\section{Laser system for triple species atom interferometry}
\begin{figure}[H]
\centering
\includegraphics[width=8cm]{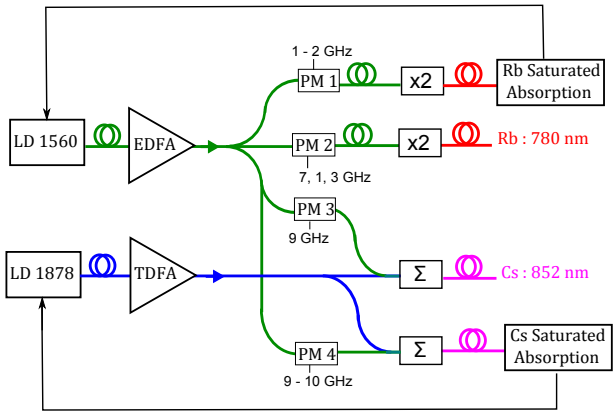}
\caption{Proposed laser architecture for \textsuperscript{85}Rb, \textsuperscript{87}Rb and \textsuperscript{133}Cs atom interferometry (LD, laser diode; EDFA, erbium-doped
fiber amplifier; TDFA, thulium-doped fiber amplifier; x2, second harmonic generation in a PPLN waveguide; $\Sigma$, sum frequency generation in a PPLN waveguide; PM, phase modulator).}
\label{triple}
\end{figure}

In the last part of this article, we propose a laser architecture (see Fig. \ref{triple}) that allows to generate all the laser frequencies needed for triple species atom interferometry with \textsuperscript{85}Rb, \textsuperscript{87}Rb and \textsuperscript{133}Cs. The manipulation of three species could find practical applications like inertial sensors without dead time or multi-axis inertial sensors \cite{Johnson:16}. An atom interferometer experiment needs several laser frequencies to perform the different needed steps: laser cooling, repumping, detection, Raman transition. For  \textsuperscript{87}Rb, the laser cooling frequency, the detection frequency and the first Raman frequency are obtained with a laser which has a frequency dynamically adjustable over 1 GHz around the transitions F=2 - F' (F' corresponds to an hyperfine excited state of the D2 line). This frequency control is achieved with the method described in \cite{Theron:15} and consists in generating sidebands at a frequency of 1 - 2 GHz thanks to a phase modulator (PM 1) and then by locking the sideband on a saturated absorption peak of \textsuperscript{85}Rb F=3 - F'. The laser at 780 nm addresses thus the transitions F=2 - F' of  \textsuperscript{87}Rb and its frequency can be dynamically adjusted by changing the radio frequency on PM 1. The repumper frequency and the second Raman frequency are obtained with PM 2 phase modulator which creates a sideband around 7 GHz addressing the transitions F=1 - F' \cite{Carraz:09}.  For Cs, we use the same method. The sideband at 9 - 10 GHz generated by the phase modulator PM 4 is locked on a saturated absorption peak of Cs F=3 - F' and thus the laser at 852 nm can be dynamically tuned over the transition F=4 - F' of Cs. The phase modulator PM 3 generates a sideband at 9 GHz and addresses the transitions F=3 - F' of Cs. For the generation of the laser frequencies of \textsuperscript{85}Rb, we use the method described in \cite{Bonnin:15} consisting in creating the laser frequencies of \textsuperscript{85}Rb  from the laser for \textsuperscript{87}Rb  by sending the appropriate radio frequencies on the phase modulator PM 2 i.e. 1 GHz for laser cooling, 3 GHz for repumper and second Raman laser frequency. Finally, our laser system is able to operate a triple species atom interferometry experiment with a fiber laser system containing only two laser sources. This is possible thanks to the 4 phase modulators on the 1560 nm laser.

\section{Conclusion}
In conclusion, we have reported the first demonstration of a multi-line fiber laser system addressing cesium atomic transitions at 852 nm. The laser system architecture is based on sum-frequency generation of a laser at 1560 nm and a laser at 1878 nm, taking advantage of the availability of well developped fibered optical components at 1.5 $\mu$m and 2 $\mu$m. Such architecture allows more particularly the laser system, composed of only two laser sources, to address all the atomic transitions for cooling and manipulating the three atomic species \textsuperscript{87}Rb, \textsuperscript{85}Rb, \textsuperscript{133}Cs. The simplicity, compactness and reliability of this laser architecture open the way to a new generation of atom sensors using Cs and Rb atoms dedicated to field applications. This multi-line laser system appears also as a promising candidate for multi-species atom interferometry and more specifically for futur UFF tests in space.

\end{multicols}

\end{document}